\documentclass[english,aps,preprint]{revtex4}
\usepackage[T1]{fontenc}
\usepackage[latin9]{inputenc}
\usepackage{color}
\usepackage{amsmath}
\usepackage{graphicx}

\makeatletter

\providecommand{\tabularnewline}{\\}

\@ifundefined{textcolor}{}
{%
 \definecolor{BLACK}{gray}{0}
 \definecolor{WHITE}{gray}{1}
 \definecolor{RED}{rgb}{1,0,0}
 \definecolor{GREEN}{rgb}{0,1,0}
 \definecolor{BLUE}{rgb}{0,0,1}
 \definecolor{CYAN}{cmyk}{1,0,0,0}
 \definecolor{MAGENTA}{cmyk}{0,1,0,0}
 \definecolor{YELLOW}{cmyk}{0,0,1,0}
}

\makeatother

\usepackage{babel}
\begin{document}

\preprint{\textcolor{black}{This line only printed with preprint option}}

\title{\textcolor{black}{\normalsize{Antiferromagnetic ground state with
pair-checkboard order in FeSe}}}

\author{\textcolor{black}{\normalsize{Hai-Yuan Cao, Shiyou Chen, Hongjun
Xiang, Xin-Gao Gong}}}

\affiliation{\textcolor{black}{Key Laboratory for Computational Physical Sciences
(MOE), State Key Laboratory of Surface Physics, and Department of
Physics, Fudan University, Shanghai 200433, China}}
\begin{abstract}
\textcolor{black}{\normalsize{Monolayer FeSe thin film grown on SrTiO$_{3}$(001)
(STO) shows the sign of T$_{c}$ > 77 K, which is higher than the
T$_{c}$-record of 56 K for the bulk FeAs-based superconductors. However,
little is known about the magnetic ground state of FeSe, which should
be closely related to its unusual superconductivity. Previous studies
presume the collinear stripe antiferromagnetic (AFM) state as the
ground state of FeSe, same to that in FeAs superconductors. Here we
find a novel magnetic order named ``pair-checkboard AFM'' as the
magnetic ground state of tetragonal FeSe. The novel pair-checkboard
order results from the interplay between the nearest, the next-nearest
and the unnegligible next-next-nearest neighbor magnetic exchange
couplings of Fe atoms. The monolayer FeSe in pair-checkbord order
shows an unexpected insulating behavior with a Dirac-cone-like band
structure related to the specific orbital order of d$_{xz}$ and d$_{yz}$
characters of Fe atoms, which could explain recently observed insulator-superconductor
transition. The present results cast new insights on the magnetic
ordering in FeSe monolayer and its derived superconductors.}}{\normalsize \par}

\textcolor{black}{PACS number: 74.25.Ha, 75.70.Ak, 74.70.-b, 73.22.-f }
\end{abstract}
\maketitle
\textcolor{black}{The high temperature (high-T$_{c}$) superconductivity
discovered in the iron-based superconductors \cite{key-1,key-2,key-4}
breaks the conventional knowledge that the magnetic atoms like Fe
should not contribute to the superconductivity. This inspires that
the magnetism plays an important role in the mechanism of the high-T$_{c}$
superconductivity in iron-based superconductors \cite{key-5}. Although
the electronic properties for different families of iron-based superconductors
can be somehow different \cite{key-6}, they all are believed to share
the common feature of AFM ordered parent compound \cite{key-7}.}

\textcolor{black}{While the magnetism contributing to the high-Tc
superconductivity has attracted wide attention \cite{key-9}, the
magnetic ground states for the parent compounds of iron-based superconductors
remain unclear. Recently, the sign of over 77 K unconventional high
T$_{c}$ superconductivity \cite{key-10,key-11,key-12,key-13} has
been observed in monolayer FeSe grown on STO substrate \cite{key-14,key-15,key-16,key-17,key-18,key-19},
which is much higher than the highest T$_{c}$ record in the intensively
studied FeAs systems \cite{key-20,key-21}. For FeAs-based materials,
the collinear AFM (or the stripe AFM) has been verified as the ground
state for the parent compounds by neutron scattering \cite{key-9}.
However, the ground state for the compound based on FeSe is still
waiting to be clarified. Previous theoretical studies presumed that
FeSe has the same ground state as FeAs-based materials \cite{key-22,key-23,key-24,key-25,key-26}.
From the previous experimental results, the electronic properties
of FeSe-based materials are much different from that of FeAs-based
materials \cite{key-27}, especially for that in monolayer FeSe on
STO substrate \cite{key-10,key-11,key-12,key-13,key-14,key-15,key-16,key-17,key-18,key-19}.
Recent ARPES experiment observed that the insulator-superconductor
transition via doping in the monolayer FeSe grown on the STO substrate
\cite{key-19}, which indicates that the ground state of monolayer
FeSe could be insulating. Besides, in the recently discovered molecular-intercalated
iron-selenide Li$_{x}$(ND$_{2}$)$_{y}$(ND$_{3}$)$_{1-y}$Fe$_{2}$Se$_{2}$
\cite{key-42}, the neutron-inelastic-scattering measurements found
that the magnetic scattering in momentum space is unusually closer
to wave-vector ($\pi$,$\dfrac{\pi}{2}$) \cite{key-43}, which means
that in FeSe there could be an unexpected magnetic order other than
collinear AFM. As yet, unveiling the ground-state properties of FeSe
would be crucial for understanding the mechanism of novel high-Tc
superconductivity.}

\textcolor{black}{Here, based on the first-principles calculations,
we find a novel magnetic order, named as the pair-checkboard AFM,
which is quite different from previously proposed magnetic order in
iron based compounds, to be the magnetic ground state (Fig. 1(a))
for tetragonal FeSe \cite{key-44}. The underlying physics of this
new magnetic order could be effectively described by a Heisenberg
model including the nearest, the next-nearest and the next-next-nearest
neighbor superexchange interaction mediated by Se-4p orbitals. This
novel ground state is metallic in the bulk FeSe, while it has a Dirac-cone-like
band structure with a non-zero band gap induced by the spin-orbit
interaction in monolayer FeSe, where the interlayer interaction between
the FeSe layers is absent. The pair-checkboard AFM would also induce
a $2\times1$ reconstruction in FeSe due to the different distance
of the Se atoms to the Fe plane. Furthermore, we confirm that this
novel pair-checkboard AFM not only exists in tetragonal FeSe but also
in its derived compounds, like the recently synthesized bulk LiFeO$_{2}$Fe$_{2}$Se$_{2}$
\cite{key-29}. This is the first time to reveal that the magnetic
ground state of FeSe is different from that of FeAs-based materials,
which could well explain several intriguing experimental observations.}

\textcolor{black}{In this letter, we performed extensive study on
the electronic and magnetic properties of FeSe based on the first-principles
simulations. We employed the plane-wave basis set and the projected
augmented wave method \cite{key-30,key-31} which is implemented in
the VASP code \cite{key-32,key-33} to calculate the electronic and
magnetic properties. We adopted the generalized gradient approximation
(GGA) with Perdew-Burke-Ernzerhof (PBE) formula \cite{key-34} for
the exchange-correlation functional. A plane-wave cutoff energy of
450 eV and a Monkhorst-Pack mesh of $16\times8$ $k$-points \cite{key-35}
for monolayer FeSe and $18\text{\ensuremath{\times}}9\times9$ $k$-points
for bulk FeSe with 0.1 eV Gaussian smearing were used in magnetic
unit cell calculations. A supercell of 16 Fe atoms was used to calculate
the magnetic exchange coupling parameters. The vacuum layer more than
20 \AA{} thick was used in the calculation for monolayer FeSe to ensure
decoupling between neighboring FeSe layers. For structural relaxation,
all the atoms were allowed to relax until atomic forces are smaller
than $0.01$ $eV/\text{\AA}$. The density of states calculations
were performed based on the tetrahedral method \cite{key-36} with
a much denser k-grid of $24\times12$ and $24\times12\times12$ for
the magnetic unit cell of monolayer FeSe and bulk FeSe, respectively.
Both the lattice constant and the atomic positions are fully optimized. }

\textcolor{black}{Our results show that the magnetic ground state
of FeSe should be the pair-checkboard AFM, rather than the collinear
AFM which was assumed to be. We have calculated the relative energy
of various magnetic orders, including the nonmagnetic state, the checkboard
AFM order (or the N�el AFM), the bicollinear AFM order, the collinear
AFM order and the pair-checkboard AFM. Table I lists the energies
and the magnetic moments for both the bulk FeSe and monolayer FeSe.
One can clearly see that the energy of pair-checkboard AFM is lower
than that of the collinear AFM order by 15 meV/f.u. and 12 meV/f.u.
for the bulk FeSe and monolayer FeSe, respectively, while the magnetic
moments in the bulk FeSe and monolayer FeSe are quite similar}\textcolor{red}{.}\textcolor{black}{{}
The breaking of the C$_{4}$ symmetry after lattice-constant optimization
is found in the pair-checkboard AFM order as well as in the collinear
AFM, which is caused by the ferro-orbital order of Fe atoms in FeSe
\cite{key-37}. }

\textcolor{black}{The pair-checkboard AFM is different from the collinear
AFM order that each spin of Fe atom has one neighbor spin aligned
ferromagnetically while the other three neighbor spins all aligned
antiferromagnetically (Fig. 1a). More interestingly, we find apparent
difference between the charge-density distribution of the pair-checkboard
AFM and the collinear AFM (Fig. 1b, 1c). Both Fe and Se atoms in the
pair-checkboard AFM order has a unique orbital order which does not
appear in other AFM orders. The Se atoms could be divided into two
groups (labeled as Se1 and Se2 in Fig. 1(b)) with the different distances
between the Se atoms and the Fe plane ($z_{Se}$). The $z_{Se}$ for
Se1 is $1.48$$\text{\AA}$ and that for Se2 is $1.46$ $\text{\AA}$
which can be regarded as a $2\times1$ reconstruction in FeSe.}

\textcolor{black}{The electronic band structure and the projected
PDOS of bulk and monolayer FeSe with the pair-checkboard AFM state
is shown in Fig. 2 and Fig. 3, respectively. For bulk FeSe, the band
structure shows that there are two electron pokets locating around
$\Gamma$ point and along $X'-\mbox{\ensuremath{\Gamma}}$ line, and
one hole pocket centered around $Z$ point. From the projected PDOS,
we find the band crossing the Fermi level mainly coming from the $d_{xz}$
and $d_{yz}$ orbitals, which is caused by the interlayer interaction
between adjacent FeSe layers in bulk FeSe. The PDOS of Fe atoms near
the Fermi level looks similar for both bulk FeSe and monolayer FeSe,
which means that the intralayer interaction in FeSe layers could probably
dominate the major physics related to the superconductivity in FeSe. }

\textcolor{black}{The tetragonal FeSe with pair-checkboard AFM order
has a feature of exotic Dirac-cone-like band structure as shown in
Fig. 2(a) and 2(b), respectively. If including the spin-orbit coupling,
monolayer FeSe with the pair-checkboard AFM becomes insulating with
a band gap around 27 meV, while that with other magnetic orders are
all metallic. It is worth to note that both the valence band maximum
(VBM) and the conduction band minimum (CBM) of the band structure
locating around the k-point (0.2, 0.0, 0.0) in $\Gamma-X$ boundary
of the first magnetic Brillouin zone. }

\textcolor{black}{The Dirac-cone-like band structure relates to a
specific orbital order of $d_{xz}$ and $d_{yz}$ characters of Fe
atoms, which can be seen from the decomposed band structure near the
Fermi level (see Supplemental Material Figure S2 \cite{key-3}). The
spin-majority $d$-orbitals of Fe atoms are almost all filled, while
the density of states (DOS) near the Fermi level is mostly contributed
by the spin-minority $d$-orbitals of the Fe atoms. For the spin-minority
part of Fe atoms, $d_{x^{2}-y^{2}}$, $d_{z^{2}}$ and $d_{xz}$ orbitals
are mostly filled while $d_{yz}$ and $d_{xy}$ orbitals are slightly
filled (Fig. 3(b)). The two bands cross over the Fermi level are mainly
composed of spin-minority $d_{xz}$ and $d_{yz}$ characters of Fe
atoms, respectively. If without the spin-orbit coupling, $d_{xz}$
and $d_{yz}$ belong to different symmetry-groups which allows them
to cross without hybridization at the Fermi level. Then turning on
the spin-orbit coupling could lead to the mixing between $d_{xz}$
and $d_{yz}$ orbitals, resulting a gap opening. The charge density
at the VBM is mainly composed of $d_{xz}$ hybridized with $d_{xy}$
and $d_{z^{2}}$ of Fe atoms, while that at the CBM is mainly composed
of $d_{yz}$ hybridized with $d_{xy}$ (Fig. 3(b)). The charge-density
distribution at the VBM and CBM is in agreement with the ferro-orbital
order of $d_{xz}$ and $d_{yz}$ orbitals\cite{key-37}. The emerging
Dirac-cone-like band structure in the pair-checkboard AFM should be
directly related to the novel magnetic ground state in the monolayer
FeSe, which means that the orbital order and the magnetic order are
strongly coupled together.}

\textcolor{black}{To further describe the origin of the new magnetic
order in FeSe quatitatively, we propose a frustrated Heisenberg model
with the nearest, the next-nearest and the next-next-nearest neighbor
couplings $J_{1}$, $J_{2}$ and $J_{3}$ \cite{key-25}.}

\textcolor{black}{
\begin{equation}
H=J_{1}\sum_{<ij>}\vec{S_{i}}\cdot\vec{S_{j}}+J_{2}\sum_{<<ij>>}\vec{S_{i}}\cdot\vec{S_{j}}+J_{3}\sum_{<<<ij>>>}\vec{S_{i}}\cdot\vec{S_{j}},
\end{equation}
whereas $<ij>$, $<<ij>>$ and $<<<ij>>>$ denote the summation over
the nearest, the next-nearest and the next-next-nearest neighbors,
respectively. The present model includes the next-next-nearest magnetic
coupling $J_{3}$, and this is crucial to correctly describe the new
magnetic ground state in FeSe. According to the band structure of
FeSe, we believe that our Heisenberg model could capture the substantial
physics in the magnetic properties. Mapping from the calculated energy
of different mangetic orders, we find that for bulk FeSe $J_{1}=47$
meV/S$^{2}$ , $J_{2}$ = 27 meV/S$^{2}$, and $J_{3}$= 7 meV/S$^{2}$
, while for monolayer FeSe $J_{1}=44$ meV/S$^{2}$ , $J_{2}$ = 25
meV/S$^{2}$, and $J_{3}$= 6 meV/S$^{2}$.}

\textcolor{black}{It is known that when $J_{2}>\dfrac{J_{1}}{2}$,
the collinear AFM rather than the checkboard AFM would be the magnetic
ground state in a $J_{1}-J_{2}$ AFM square lattice. If further including
$J_{3}>\dfrac{J_{2}}{2}$ in the model, the magnetic ground state
would turn out to be the bicollinear AFM \cite{key-25}. In our calculation
for FeSe, we find that $J_{2}>\dfrac{J_{1}}{2}$ and $\dfrac{J_{2}}{2}>J_{3}>\dfrac{2J_{2}-J_{1}}{2}$,
and under this condition the new magnetic order we proposed turns
out to be the magnetic ground state in FeSe.}

\textcolor{black}{LiFeO$_{2}$Fe$_{2}$Se$_{2}$, a FeSe-based superconductor
with T$_{c}$ $\sim$43 K and neutral LiFeO$_{2}$ anti-PbO-type spacer
layers intercalated between FeSe layers, was recently synthesized
by the hydrothermal method \cite{key-29}. The relative energy difference
between various magnetic orders in LiFeO$_{2}$Fe$_{2}$Se$_{2}$
is similar to that in FeSe (see Supplemental Material Table S1 \cite{key-3}).
The calculation shows that the pair-checkboard AFM is not only the
magnetic ground state for LiFeO$_{2}$Fe$_{2}$Se$_{2}$, but its
relative stability to other magnetic states is even more robust comparing
to that of monolayer FeSe. This reveals that the pair-checkboard AFM
could be the universal magnetic ground state for FeSe layer and its
derived undoped materials. Besides that, we also find that the Dirac-cone-like
band structure still be kept in the bulk LiFeO$_{2}$Fe$_{2}$Se$_{2}$.}

\textcolor{black}{In previous experiments, there was no direct observation
for the pair-checkboard AFM order in bulk tetragonal FeSe. This could
probably be attributed to the absense of the high quality FeSe sample.
The extra Fe atoms in the FeSe samples could suppress the stability
of the pair-checkboard AFM order (see Supplemental Material Figure
S4\cite{key-3}). Although the pair-checkboard AFM order has not been
observed directly yet, there were several experimental evidences indicating
the existence of this novel magnetic order. At first, the 27 meV band
gap we predicted in FeSe monolayer with pair-checkboard AFM order
is quite closed to the ARPES experimentally observed insulating gap
of 20\textasciitilde{}25 meV \cite{key-19}, while that with other
magnetic orders all remains to be metallic. Another STM experiment
observed a $2\times1$ reconstruction would occur in monolayer FeSe
on the STO substrate \cite{key-40}, which is also coincident with
two types of Se atoms with different height $z_{Se}$ in pair-checkboard
AFM. Very recently, electric transport measurements observed the Dirac-cone-like
ultrafast carriers in the single crystal FeSe superconductor \cite{key-41},
which could also be interpreted as originating from our proposed pair-checkboard
AFM order induced Dirac-cone-like band structure.}

\textcolor{black}{In summary, the present studies reveal that FeSe
does not share the same collinear AFM magnetic ground state with the
FeAs-based materials. The magnetic ground state of FeSe is pair-checkboard
AFM, which is metallic in bulk FeSe and insulating with a 27 meV band
gap in monolayer FeSe and completely different from the magnetic states
found in other iron-based superconductors. Such novel magnetic order
is found to be robust against tensile strain up to a few percent and
also robust against electron doping to a certain level (see Supplemental
Material Figure S4 \cite{key-3}). The properties of the predicted
gapped insulating ground state are in good agreement with the recent
experimental observations \cite{key-19,key-40,key-41}. The pair-checkboard
AFM order in FeSe shed new lights on the understanding of high-T$_{c}$
superconductivity in FeSe monolayer on the oxides substrates and FeSe-layer
derived superconductors. The novel pair-checkboard AFM order we predicted
would call more direct experiments for investigating the magnetic
properties in high-quality FeSe samples.}
\begin{acknowledgments}
\textcolor{black}{We acknowledge professor D. L. Feng, Dr. Rui Peng
and Dr. Shiyong Tan for stimulating discussions. The work was partially
supported by the Special Funds for Major State Basic Research, National
Natural Science Foundation of China (NSFC), Program for Professor
of Special Appointment (Eastern Scholar) and the National Basic Research
Program of China (973 Program). Computation was performed in the Supercomputer
Center of Fudan University. }

\pagebreak{}
\end{acknowledgments}
\textcolor{black}{}
\begin{figure}
\textcolor{black}{\includegraphics[scale=0.5]{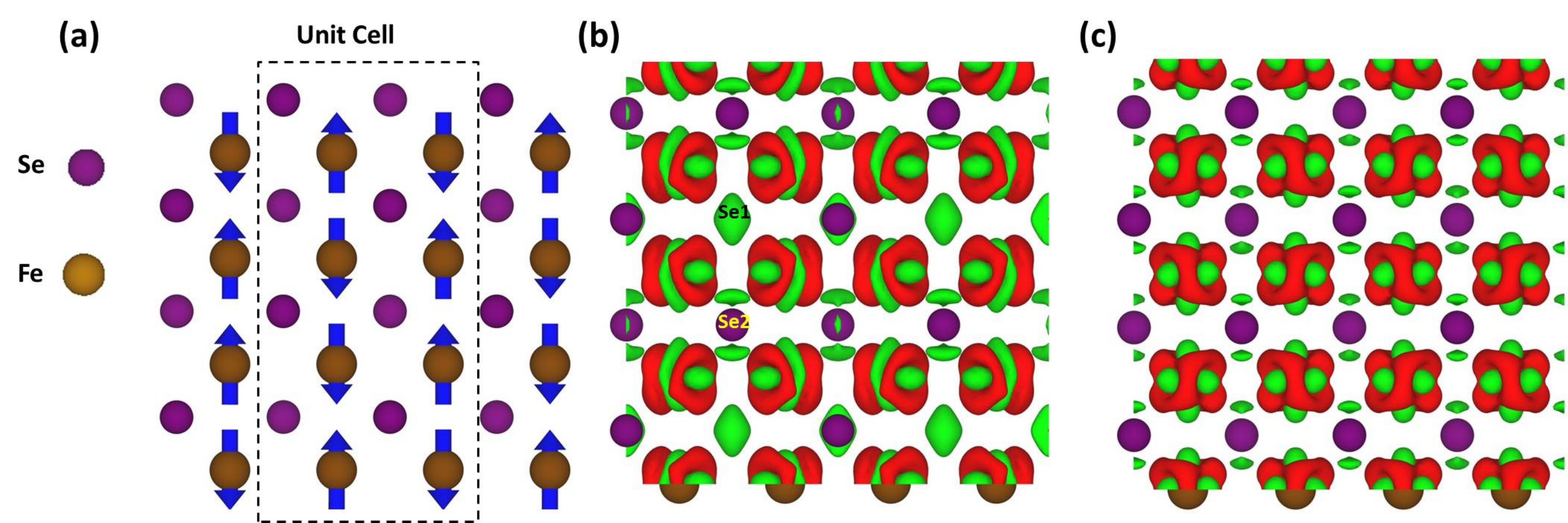}}

\centering{}\textcolor{black}{\caption{(a) Schematic top view of the pair-checkboard AFM in FeSe layer. Each
spin of Fe atom has only one neighbor spin aligned ferromagnetically
while the other neighbor spins all aligned antiferromagnetically.
The rectangle enclosed by the dashed lines denotes the magnetic unit
cell. (b) The charge density difference between the pair-checkboard
AFM and the nonmagnetic state. (c) The charge density difference between
the collinear AFM and the nonmagnetic state. The isosurface depicted
by the red and green colors represents the lost and gained charge
density comparing to the nonmagnetic state. Se atoms in the pair-checkboard
AFM order show an orbital order which are labeled with Se1 (blue)
and Se2 (yellow), respectively.}
}
\end{figure}

\textcolor{black}{}
\begin{table}
\begin{centering}
\textcolor{black}{}%
\begin{tabular}{|c|c|c|c|c|}
\hline 
 & \textcolor{black}{$\Delta E$$_{Bulk}$ (meV/f.u.)} & \textcolor{black}{M$_{Bulk}$ ($\mu_{B}$)} & \textcolor{black}{$\Delta E$$_{Mono}$ (meV/f.u.)} & \textcolor{black}{M$_{Mono}$ ($\mu_{B}$)}\tabularnewline
\hline 
\hline 
\textcolor{black}{Nomangetic Order} & 0 & \textcolor{black}{0} & \textcolor{black}{0} & \textcolor{black}{0}\tabularnewline
\hline 
\textcolor{black}{Collinear AFM} & -69 & 1.9 & \textcolor{black}{-87} & \textcolor{black}{1.9}\tabularnewline
\hline 
\textcolor{black}{Checkboard AFM} & -41 & 1.7 & \textcolor{black}{-62} & \textcolor{black}{1.7}\tabularnewline
\hline 
\textcolor{black}{Bicollinear AFM} & -21 & 2.2 & \textcolor{black}{-37} & \textcolor{black}{2.2}\tabularnewline
\hline 
\textcolor{black}{pair-checkboard AFM} & -84 & 2.0 & \textcolor{black}{-99} & \textcolor{black}{2.0}\tabularnewline
\hline 
\end{tabular}
\par\end{centering}

\centering{}\textcolor{black}{\caption{Energy difference $\Delta E$ (reference to the energy of the nonmagnetic
state) and magnetic moments $M_{Fe}$ of the bulk and monolayer FeSe
with different magnetic order. }
}
\end{table}

\textcolor{black}{}
\begin{figure}
\begin{centering}
\textcolor{black}{\includegraphics[scale=0.6]{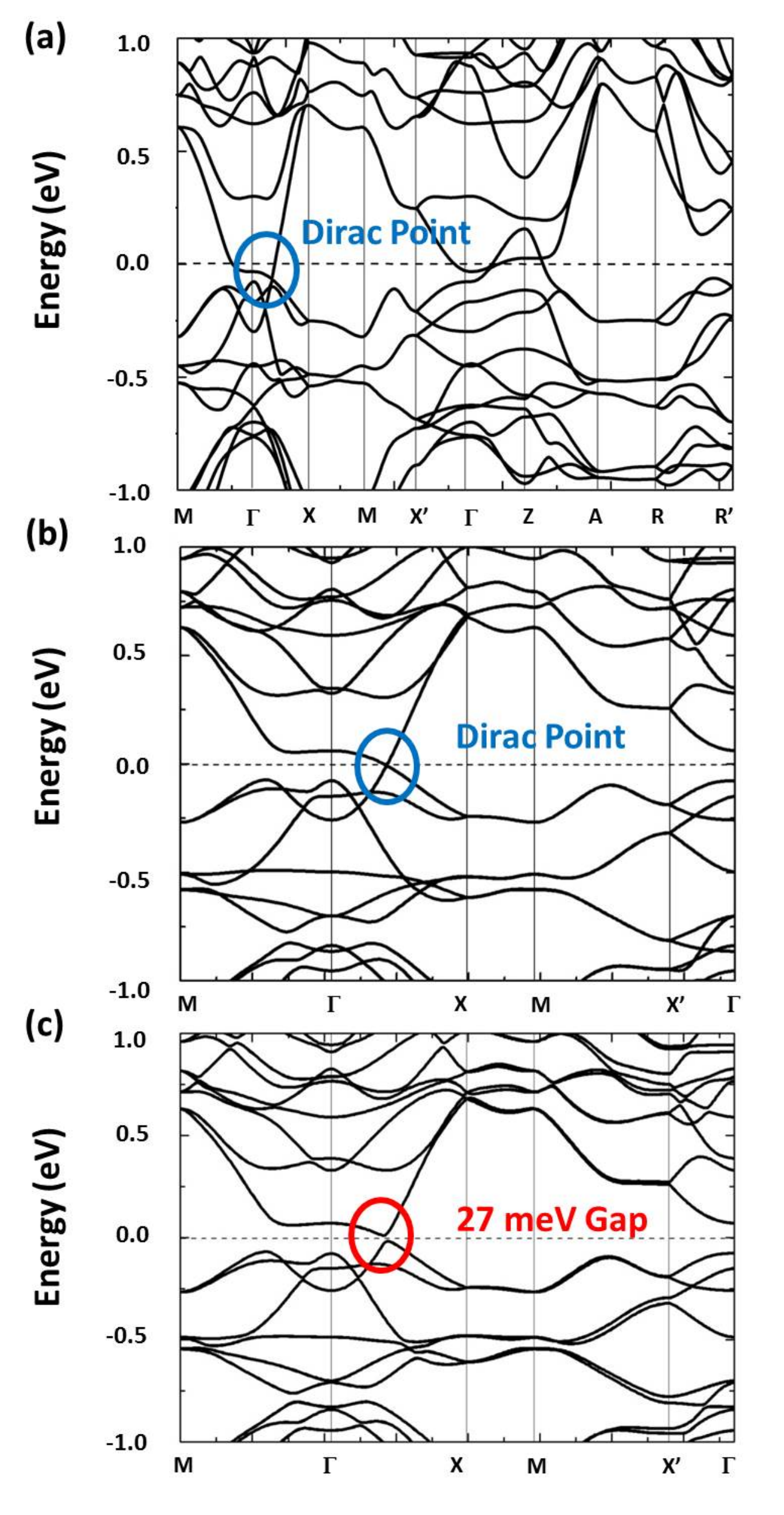}}
\par\end{centering}

\centering{}\textcolor{black}{\caption{(a) Electronic band structure of bulk FeSe with the pair-checkboard
AFM. (b) and (c) Electronic band structure of monolayer FeSe in the
pair-checkboard AFM order without or with the spin-orbit coupling.
The Dirac-point appears in the bulk FeSe and monolayer FeSe (highlighted),
while including SOC would open (highlighted) a 27 meV indirect band
gap in monolayer FeSe. The Fermi level is denoted by the dashed line.}
}
\end{figure}

\textcolor{black}{}
\begin{figure}
\begin{centering}
\textcolor{black}{\includegraphics[scale=0.6]{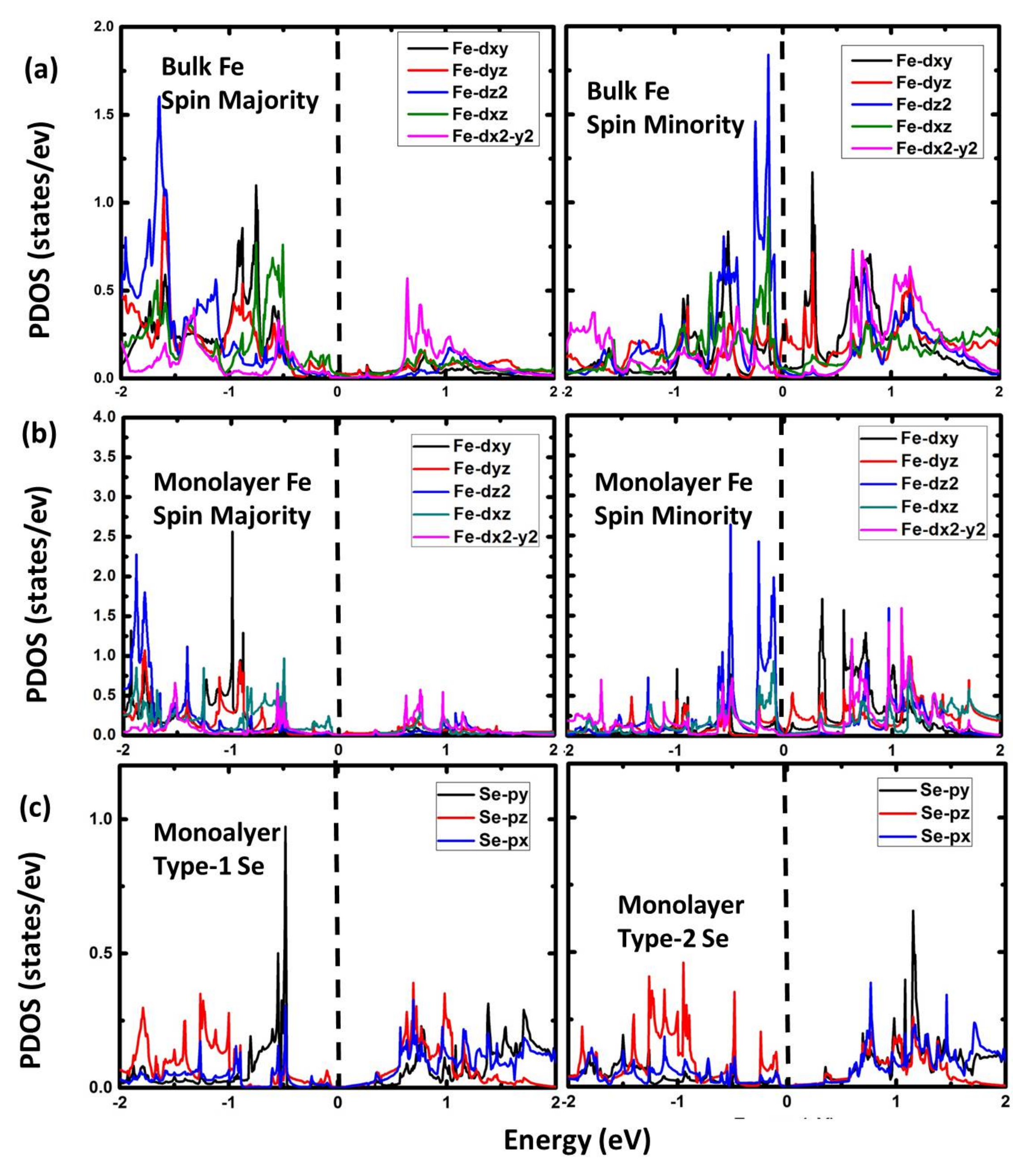}}
\par\end{centering}

\centering{}\textcolor{black}{\caption{Projected orbital-resolved partial density of states of (a) spin-majority
part and spin-minority part of Fe atom in bulk FeSe, (b) spin-majority
part and spin-minority part of Fe atom in monolayer FeSe, (c) type-1
Se atom and type-2 Se atom (defined in Fig. 3(b)), respectively. \textcolor{black}{The
d-orbitals }of Fe atoms diminish to 0 at the Fermi level in monolayer
FeSe, while the\textcolor{black}{{} spin-minority part of $d_{xz}$
and $d_{yz}$ orbitals }cross the Fermi level in bulk FeSe. The position
of the Fermi level is denoted by the dashed lines.}
}
\end{figure}

\clearpage{}

\end{document}